\documentclass[journal=nalefd,manuscript=article,layout=twocolumn,varwidth]{achemso}

\usepackage[version=3]{mhchem} 
\usepackage{latexsym} \usepackage{amsmath} \usepackage{dcolumn}
\usepackage{bm} \usepackage{graphicx} \usepackage{epsfig}
\usepackage{float} \usepackage{stackrel,amssymb}
\usepackage{multirow}

\usepackage[]{graphicx}
\usepackage[]{wrapfig}
\usepackage[]{color}

\author{Towfiq Ahmed}
\email{atowfiq@lanl.gov}
\affiliation
{Theoretical Division,
Los Alamos National Laboratory, Los
Alamos, New Mexico 87545}
\author{N. A. Modine}
\affiliation
{Center for Integrated Nanotechnologies, Sandia
National Laboratories, Albuquerque, New Mexico 87185}
\author{Jian-Xin Zhu}
\email{jxzhu@lanl.gov}
\affiliation
{Center for Integrated Nanotechnologies, Los Alamos
National Laboratory, Los Alamos, New Mexico 87545}
\altaffiliation
{Theoretical Division,
Los Alamos National Laboratory, Los
Alamos, New Mexico 87545}

\title[\texttt{achemso}]
{
Graphene/MoS$_2$ van der Waals Bilayer as the Anode Material for Next Generation Li-ion Battery: A First-Principles Investigation
}

\keywords{Density functional theory, Li ion battery, van der Waals.
 }

\begin{document}

\begin{abstract}
We performed density functional theory (DFT) calculations for a bi-layered heterostructure
combining a
graphene layer with a MoS$_2$ layer with and without intercalated Li atoms.
Our calculations demonstrate the importance of the van der Waals (vdW) interaction, which
is crucial for forming stable
bonding between the layers.
Our DFT calculation correctly reproduces the
linear dispersion, or Dirac cone, feature at the Fermi energy for the isolated graphene monolayer and the band gap for the
MoS$_2$ monolayer. For the combined graphene/MoS$_2$ bi-layer,
we observe interesting electronic structure and density of
states (DOS) characteristics near
the Fermi energy, showing both the gap like features of the MoS$_2$ layer and in-gap states with linear dispersion contributed mostly by the graphene layer.
Our calculated total
density of states
(DOS) in this vdW heterostructure reveals that
the graphene layer significantly contributes
to pinning the Fermi energy at the center of the band gap of MoS$_2$. We also find that intercalating Li ions in between the layers of the graphene/MoS$_2$ heterostructure enhances
the binding energy through orbital hybridizations between cations (Li adatoms)
and anions (graphene and MoS$_2$ monolayers). Moreover, we calculate
the dielectric function of the Li intercalated graphene/MoS$_2$ heterostructure, the imaginary component of which can be directly compared with
experimental
measurements of optical conductivity
in order to validate our theoretical prediction.
We observe sharp features in the imaginary component of the dielectric function, which shows the presence of a Drude peak in the optical conductivity, and therefore metallicity in
the lithiated graphene/MoS$_2$ heterostructure.
\end{abstract}


\section{Introduction}
\begin{figure}[!ht]
\includegraphics[width=3.5in]{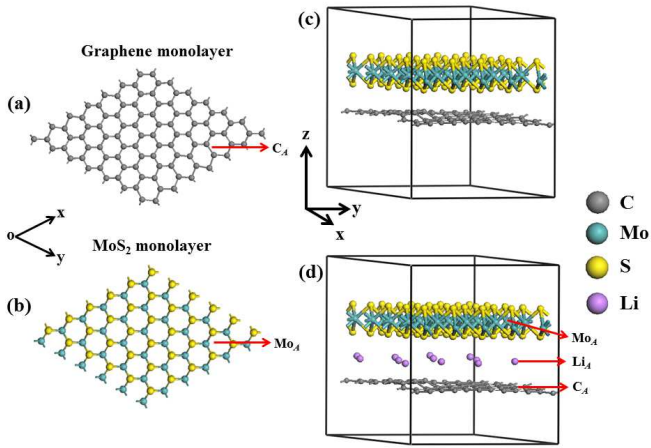}
\caption{\small Image of {\bf {(a)}} graphene monolayer,
{\bf {(b)}} MoS$_2$ monolayer,
{\bf {(c)}} unit cell for graphene/MoS$_2$ bilayer, and
{\bf {(d)}} unit cell for Li intercalated graphene/MoS$_2$ bilayer; For a
particular Li atom Li$_A$, the nearest Mo and C atoms are denoted with
Mo$_A$ and C$_A$;
}
\label{f1}
\end{figure}

Currently, Li-ion batteries (LIB) are a very popular power
source for portable electronic devices and hybrid vehicles.
However, despite being renewable and environmentally friendly, conventional Li-ion batteries are
still far from meeting the national US ABC target for a cheap (\$150/KWh) and
high capacity (3000 KWh/Kg) power source.\cite{gaines}
This is mostly due to the widespread use of graphite based anodes in LIB, which only have a small theoretical specific capacity (372 mAh/g).\cite{doeff}
Thus, the development of the next generation LIB calls for identifying alternative materials for the negative and positive electrodes. In this respect, both graphene
and MoS$_2$ monolayers have demonstrated potential as materials
for designing superior electrodes with high specific capacity. High capacities from
600 to 1000 mAh/g has been observed for single atomically thin graphene nanosheets.\cite{wang,chang,graff,lian}
Moreover, several experimental works\cite{dai,ren} on transition metals and metal oxides supported on
graphene as negative electrodes have shown enhanced performance for LIB. On the other hand, the usage of a transition metal sulfide MoS$_2$, with graphene like structure, was
proposed as a potential anode material in a patent in 1980.\cite{brandt}
Despite its high insertion capacity ($\approx$ 1000 mAh/g), MoS$_2$ alone has not been realized as the primary anode material
due to its low cyclic stability.\cite{chang}

In this Letter, we construct van der Waals layers\cite{geim} as the `intercalating compound'\cite{chang}
using graphene and MoS$_2$ bilayer.
Our density functional theory (DFT) calculations show orbital hybridization between the layers, where the vdW interaction
dominantly contributes to the stability of the heterostructure with significant binding energy.
A shift in chemical potential is also observed in our
calculations for the graphene/MoS2 bilayer when compared with the
isolated graphene or MoS$_2$
mono layers.
Particularly noticeable is the linear dispersion in the DOS near the Fermi
energy of the graphene/MoS$_2$
bilayer, which resembles that of the isolated graphene monolayer.
Thus, combining graphene with a MoS$_2$ layer helps to
pin the chemical potential near the gap center of MoS$_2$ and creates
semi-metal like electronic structure in the heterostructure.

Our DFT calculation shows an even larger binding energy when the graphene/MoS$_2$
bilayer are intercalated with Li ions. The
higher stability in such heterostructure makes it a potential candidate
for designing LIB anodes
with larger specific capacity and longer
cycle life-time. The hybridization between carbon $p$, lithium $s$, and molybdenum $d$ orbitals causes the chemical potential
to shift into the $d$-bands making the Li intercalated
heterostructure fully metallic. The resulting enhancement of the
conductivity in lithiated graphene/MoS$_2$ heterostructure is highly
desirable for the anode materials in LIB.

Earlier theoretical works\cite{cohen} have demonstrated similar
chemical potential shift using various adatoms in addition to
Li, for single and bilayer graphene.
For a related graphene and MoS$_2$ heterostructure,
Miwa {\it {et al.}}\cite{miwa} reported the DFT
calculated density of states without observing any Fermi energy pinning or linear dispersion near the gap center of
MoS$_2$. In this Letter, we have  demonstrated, for the first time, the interesting
Dirac dispersion characteristics in the electronic DOS
of graphene/MoS$_2$ bilayer at the Fermi energy. Although the lithiation makes
the compound metallic as expected,
discharging of LIB gradually turns the anode into semimetal while retaining the stability. In the language of electrochemistry, such processes can be presented  by the following two half-reactions:\cite{gold}

In anode:

\(xLiC_6/MoS_2 \stackrel[discharge]{charge}{\leftrightarrows} xLi^+ + xe^- + C_6/MoS_2 \),

and in cathode:

\(Li_{1-x}CoO_2 + xLi^+ + xe^-  \stackrel[discharge]{charge}{\leftrightarrows} LiCoO_2 \)

Therefore, our proof-of-principles calculations demonstrate
the Li intercalated graphene/MoS$_2$ bilayer
as potential candidates for designing high performance anode materials for the
next generation LIBs.

\begin{table*}
\caption{Binding Energy of graphene and MoS$_2$ interface with and without Li intercalation using DFT+vdW}
\centering
\begin{tabular}{c||c c c c c }
\hline
  & Graphene & MoS$_2$ & Graphene + & Graphene /   &  Graphene /   \\[0.5ex]
  &          &         &    Li      &    MoS$_2$   &    MoS$_2$ +     \\[0.5ex]
  &          &         &            &              &      Li \\[0.5ex]
\hline
\hline
Total Energy (eV) & -553.25348 & -414.14277 & -571.97493 & -972.07776 & -1001.27726\\[0.5ex]
             &  &  & &  &  \\[0.5ex]
             &  &  & & 0.1873 & 0.6064 \\[0.5ex]
Binding      &  &  & & (per Mo ion) & (per Mo ion) \\[0.5ex]
Energy (eV)  &  &  & &              & 1.2633 \\[0.5ex]
  &  &  & &              & (per Li ion) \\[0.5ex]
\hline
\end{tabular}
\label{table}
\end{table*}

\section{Computational Methods}

All calculations in this paper are performed using the plane-wave pseudo-potential
code VASP\cite{vasp1,vasp2,vasp3} under the generalized gradient approximation of Perdew, Burke, and Ernzerhof (PBE).\cite{pbe}
For atomic core-levels, we have used projected augmented wave (PAW)
potentials\cite{pseudo1,pseudo2} treating the
2$s$2$p$ of C, 2$s$ of Li, 4$p$5$s$4$d$ of Mo, and 3$s$3$p$ of S as the explicit valence electrons.
For all calculations, the total energy during electronic relaxation is converged to 10$^{-6}$ eV while the force/atom during
ionic relaxation is converged to 0.01eV/\AA. A maximum energy cutoff of 500 eV is used for plane-wave basis set.

In the $xy$-plane, 6x6 unit cells of graphene and 5x5 unit cells of MoS$_2$ are placed in a super-cell with $a$=$b$=16 \AA, $c$=40 \AA, $\alpha=\beta$=90$^{\circ}$, and $\gamma$=120$^{\circ}$. In our calculations, the Li atoms are intercalated between the graphene and
MoS$_2$ layers at the center of the graphene honeycomb sites. For different metal ions, this location has been shown to have
the largest binding energies by Chan {\it {et al.}}\cite{cohen}  We have placed 12 Li atoms homogeneously
in our super-cell. This sparse distribution of Li atoms approximates the interaction of isolated Li atoms with graphene and
MoS$_2$. The distance between Li atoms are large enough ($\approx$ 7 \AA) to minimize the orbital overlap between neighboring
Li atoms.

To incorporate the vdW interaction between the graphene and MoS$_2$ layers, we have used optB86b-vdW functional where the
exchange functionals were optimized for the correlation part.\cite{vdw}
Therefore, the LDA correlation part present in the
PBE functional is removed by using the parameter AGGAC = 0.000 in the input file in order to avoid double-counting.

Calculations for the isolated 6x6 graphene, 5x5 MoS$_2$, and 6x6 graphene + Li adatoms were performed using the same sized
super-cell. For ionic relaxation of graphene + Li , graphene + MoS$_2$, and graphene + Li + MoS$_2$ systems, we have used the
$\Gamma$ point to sample the Brillouin zone, while for all other calculations, e.g., DOS and dielectric functions, we
used equally distributed 80 $k$-points in the irreducible Brillouin zone.

\section{Results and Discussion}
\begin{figure*}[htpb]
    \begin{minipage}[!t]{0.68\linewidth}
    \epsfig{file=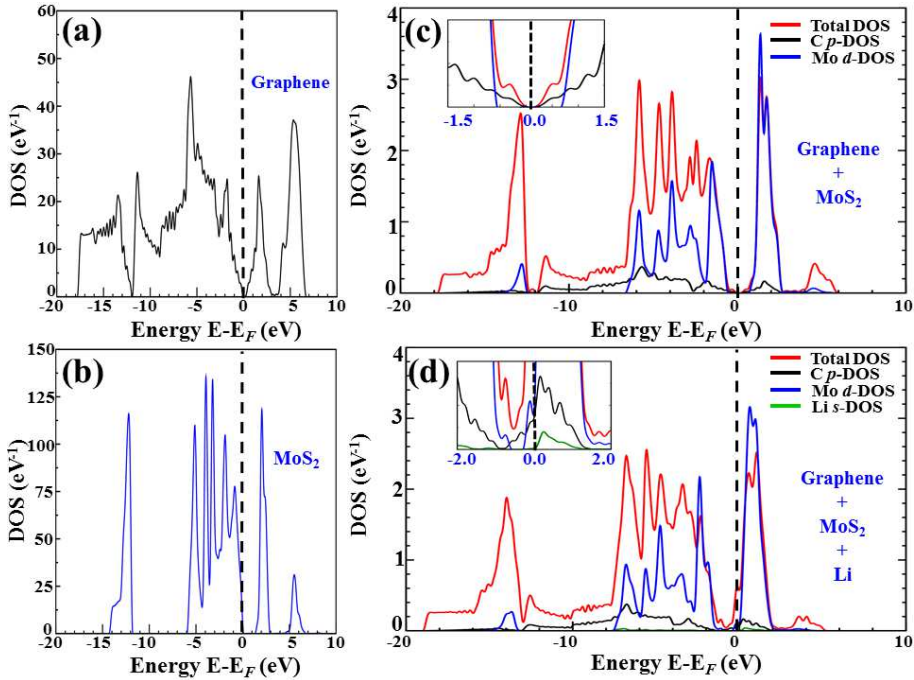, width=\linewidth}
    \end{minipage}\hfill
    \begin{minipage}[!t]{0.30\linewidth}
\caption{\small(color online)
{\bf {(a)}} Total DOS (black) for graphene;
{\bf {(b)}} Total DOS (blue) for MoS$_2$;
{\bf {(c)}} Total DOS (red) for graphene/MoS$_2$ heterostructure,
orbital projected $d$-DOS of Mo (blue), and $p$-DOS (black) for C; and
{\bf {(d)}} Total DOS (red), $d$-DOS (blue) of Mo in MoS$_2$, $p$-DOS (black)
of C in graphene, and $s$-DOS (green) of Li
in the lithiated graphene/MoS$_2$ heterostructure;
the Fermi energy is at 0 eV, shown by the vertical dashed line. Total
DOS in the unit cell (for (c) and (d)) are scaled down by a factor of 50 for
visual clarity. Inset in (c) and (d) shows DOS near the Fermi level.
}\label{f2}
\end{minipage}
\end{figure*}

To estimate the adsorption energy, we define,
\begin{equation}
\Delta E = E_A + E_B - E_{A+B},
\end{equation}
where E$_{A+B}$ is the total energy of the A+B composite system,
and E$_A$/E$_B$ is the
total energy of the  isolated A/B constituent system.

In Table 1, we present our DFT calculated total energy for graphene, MoS$_2$,
the graphene/MoS$_2$ heterostructure, and the lithiated graphene/MoS$_2$ heterostructure. Using Eq.~(1),
we calculate the binding energy between graphene and MoS$_2$ bilayer with
and without Li ions in between. We have 25 Mo atoms in our supercell, and
scaled the total adsorption energy to obtain $\Delta$E per Mo ion.
 We find the $\Delta$E per Mo ion is enhanced by about three times when Li intercalation
is considered. Without Li atoms, the graphene and MoS$_2$ layers
form a physisorbed heterostructure with $\approx$ 0.1873 eV adsorption energy
per Mo ion,
where the most important contribution comes from the surface van der Waals
interaction. The adsorption energy significantly increases in the presence of
intercalated Li ions, and these higher adsorption energies (0.6 eV per
Mo ion or 1.26 eV per Li ion) suggest a more stabile structure.  At the same time, the
small change in atomic positions during the ionic relaxation clearly indicates
the process is dominantly van der Waals driven physisorption. Thus, the Li ions
are easily separable and lithiation process can be reversed. This is a highly
desirable criterion for LIBs during ionization-deionization process which
enhances the cycle lifetime of batteries.

The electronic structure of the isolated graphene and MoS$_2$
monolayers is changed near the Fermi energy when they
are assembled to form a van der Waals heterostructure. In ~\ref{f2}, we
present the DOS for the graphene/MoS$_2$ bilayer systems with and without
lithiation and
compare with the DOS of the constituent monolayers. A graphene monolayer is a two
dimensional Dirac material with linear dispersion near the
Fermi energy.\cite{sasha,david} This property manifests by the cone like
feature in the DOS of graphene at the Fermi energy as shown in ~\ref{f2}(a).

On the other hand, MoS$_2$ monolayer is a direct gap
semiconductor with $\approx$ 1.8 eV band gap.\cite{mak} In our DFT
calculations,
this band gap is estimated to be 1.4 eV as shown in ~\ref{f2}(b).
Due to the unfilled $d$-orbitals in
Mo atoms,
the semi-local DFT functionals are inadequate, and one can incorporate many-body
corrections such as GW\cite{pan} or hybrid functionals HSE\cite{jason} which
accounts for the missing 0.4 eV band gap in the conventional DFT.
In the present proof-of-principles
calculations, our findings should remain unaffected from such band
gap correction,
while incorporating these corrections in large supercell with vdW
interaction would greatly increase the computational cost. Therefore,
we neglect such band-gap corrections in our calculations.

We plot the total DOS of graphene/MoS$_2$ in ~\ref{f2}(c) with solid red curve.
To provide visual clarity, total DOS is scaled down by a factor of 50
while comparing
 with the partial $d$-DOS (blue) of
Mo in MoS$_2$ and $p$-DOS (black) of C in graphene. We find the presence of both the
Dirac-cone-like feature from graphene and a gap-like feature from MoS$_2$ at the
Fermi energy of the composite bilayer of graphene/MoS$_2$. An enlarged
version of these symmetric features in the DOS at E$_F$ is
presented in the inset of ~\ref{f2}(c).
This finding can be interpreted as the Fermi energy pinning at the gap center
of MoS$_2$ with the Dirac states contributed from the graphene monolayer.
Such pinning of E$_F$ cannot be identified in the isolated MoS$_2$
monolayer where the
exact location of the chemical potential is arbitrary inside the gap.
The effect of lithiation on the DOS is shown in~\ref{f2}(d). Using the similar
scaling, we compare the total DOS with that of the partial $d$, $p$, and $s$
orbitals
of Mo, C, and Li atoms. The Mo and C atoms are selected from the MoS$_2$ and
graphene layers, which are closest to a particular Li atom. These atoms
are labeled in ~\ref{f1}(d).
However, we find the DOS is no longer symmetric in the presence of Li adatoms.
Due to a strong hybridization between Li $s$, Mo $d$ and C $p$
orbitals as well as the electron doping effect of Li, the chemical potential is shifted towards the unoccupied $d$ bands,
and thus, composite system becomes metallic. This implies that the loss of electrons from 
Li to the graphene and MoS$_2$ layers leads to a strong Coulomb interaction between Li ions and charged layers. It explains our observation of an increased binding energy with 
Lithiation.
\begin{figure*}[htpb]
    \begin{minipage}[!t]{0.68\linewidth}
    \epsfig{file=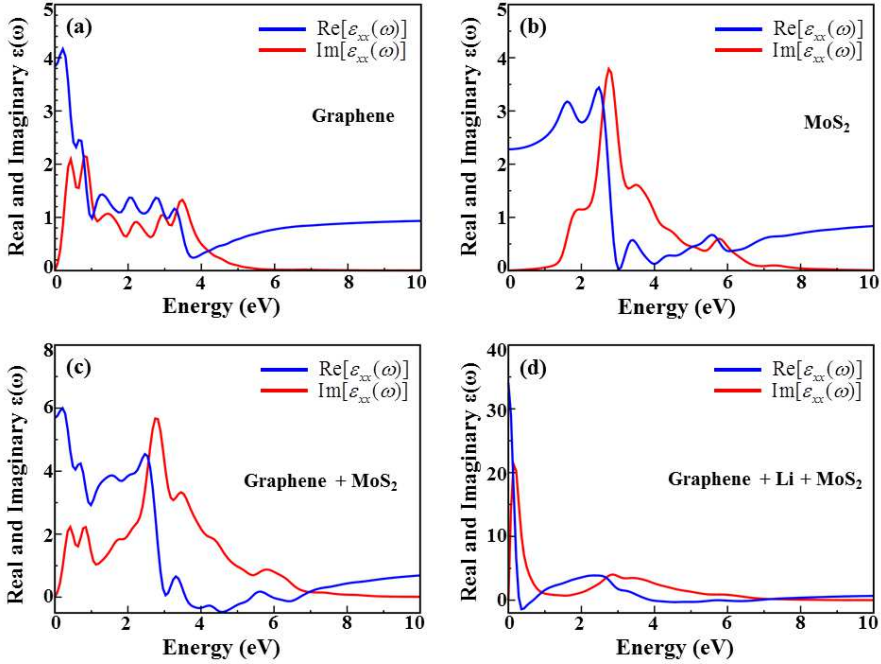, width=\linewidth}
    \end{minipage}\hfill
    \begin{minipage}[!t]{0.30\linewidth}
\caption{\small (color online)  Calculated real (blue) and imaginary (red) part of the
dielectric function for
{\bf {(a)}} graphene,
{\bf {(b)}} MoS$_2$,
{\bf {(c)}} graphene/MoS$_2$ bilayer, and
{\bf {(d)}} Li intercalated graphene/MoS$_2$ heterostructure;
Due to the proportionality between optical conductivity and
imaginary dielectric function, such results can be directly
compared with experimental measurements. See the main text for details.
}
\label{f3}
\end{minipage}
\end{figure*}

Our calculations for the real and imaginary part of dielectric function
are presented in~\ref{f3}. Imaginary dielectric function has a
direct relation to the optical conductivity via,\cite{draxl}
\begin{equation}
\textrm{Re}\,\sigma_{\alpha \beta}(\omega) = \frac{\omega}{4 \pi}\textrm{Im}\,\epsilon_{\alpha \beta}(\omega).
\end{equation}

While the conducting nature of graphene can be understood by the sharp rise
of the peak
at the zero frequency as shown in~\ref{f3}(a) (solid red curve), the gap is evident for the
isolated
MoS$_2$ monolayer as shown in~\ref{f3}(b).
When the bilayer of graphene/MoS$_2$ is formed, the total conductivity,
as can be seen in ~\ref{f3}(c), appears to be a linear combination of
contributions from each individual mono layers. On the other hand, lithiation
causes metallicity in graphene/MoS$_2$ bilayer, where a large Drude peak
appears near the zero frequency (\ref{f3}(d)).

\section{Conclusions}

In this Letter, adsorption of graphene on a MoS$_2$
monolayer is studied using DFT-based first-principles theory with and without
intercalated Li  atoms. Formation of a stable heterostructure between the layers is observed,
where the major contribution to binding comes from the vdW interaction.
our calculations show that the addition
of the graphene layer pins the Fermi energy near the gap center of MoS$_2$.
Precise identification of the Fermi energy
opens up the potential applications for LIB anode material through the
manipulation of
electronic structure of various constituent element combination with
graphene/MoS$_2$ bilayer. Lithium insertion or intercalation between
graphene/MoS$_2$ bilayer is
found to have a binding energy within the physisorption range in our calculations, which
demonstrates their prospect as superior anode materials.
Therefore, our proof-of-principles calculations of adsorption energy,
structural stability, electronic
structure, and optical properties in graphene/MoS$_2$ heterostructure
provides a basis for future experimental and theoretical study towards
designing the next generation anodes for Lithium ion batteries.

\acknowledgement
We thank Jinkyoung Yoo and Enkeleda Dervishi  for stimulating discussions.
This work was supported by U.S. DOE at Los Alamos National Laboratory
under Contract No. DE-AC52-06NA25396 and at Sandia National Laboratory
under Contract No. DE-AC04-94AL85000, and the U.S. DOE Office of Basic Energy Sciences.
The work was supported in part by the Center for Integrated
Nanotechnologies --- a U.S. DOE BES user facility.


\bibliography{references}
\end{document}